\documentclass[12pt]{article}
\usepackage[left=2.5cm,top=2.5cm,right=2.5cm,bottom=2.5cm,nohead,nofoot]{geometry}
\usepackage{graphicx}
\usepackage{hyperref}
\usepackage{setspace}
\usepackage[square]{natbib}
\usepackage{amssymb}
\usepackage{color}
\usepackage{verbatim}

\title{Exploring the Consequences of IED Deployment with a Generalized Linear Model Implementation of the Canadian Traveller Problem}
\author{Andrew C. Thomas \thanks{Visiting Assistant Professor, Department of Statistics, Carnegie Mellon University. Corresponding author; email \href{mailto:act@acthomas.ca}{act@acthomas.ca}.} \and Stephen E. Fienberg\thanks{Maurice Falk University Professor of Statistics and Social Science, Carnegie Mellon University. This work was sponsored by DARPA grant 21845-1-1130102.}}
\date{\today}

\begin{document}
\maketitle

\begin{abstract}
The deployment of improvised explosive devices (IEDs) along major roadways has been a favoured strategy of insurgents in recent war zones, both for the ability to cause damage to targets along roadways at minimal cost, but also as a means of controlling the flow of traffic and causing additional expense to opposing forces. Among other related approaches (which we discuss), the adversarial problem has an analogue in the Canadian Traveller Problem, wherein a stretch of road is blocked with some independent probability, and the state of the road is only discovered once the traveller reaches one of the intersections that bound this stretch of road.
We discuss the implementation of ideas from social network analysis, namely the notion of ``betweenness centrality'', and how this can be adapted to the notion of deployment of IEDs with the aid of Generalized Linear Models (GLMs): namely, how we can model the probability of an IED deployment in terms of the increased effort due to Canadian betweenness, how we can include expert judgement on the probability of a deployment, and how we can extend the approach to estimation and updating over several time steps.
\end{abstract}



\onehalfspacing

\section{Introduction}

Vehicles traverse a network of roads which may be compromised by an adversary with the placement of improvised explosive devices (IEDs). At a minimum, compromised roads can be avoided and replaced by alternate routes, though more typically, a convoy will spot a suspected IED while already on the road, which will require time and effort to disarm. When considering the information available to the drivers, the routers and the adversary, this can lead to a complex game-theoretic scenario. When modelling the placement of an IED as a stochastic process, so that the adversary places IEDs without regard to the response of the protagonists, we can set the problem in a decision-theoretic framework, as in \citet{singpurwalla2008nrde, singpurwalla2008einrde}.

A somewhat more systematic and technical description of the problems at hand has the following components:

\begin{itemize}

\item We have a pre-existing road system with known travel times and capacities, or these can at least be approximated. This road system represents a network at its highest capacity.

\item An adversary can interfere with this road system; this interference effects changes in the traversable graph, possibly (likely) in the middle of the traversal process.

\item We have a single convoy of vehicles (possibly a single vehicle) travelling from one node to another on the graph. This convoy can be divided into multiple sub-units, which may take separate paths, with some additional cost for security but some additional benefit for increased probability of  ``success.'' 

\item The goal is to minimize some objective function based on the time of travel, expenses for protection and actual transport (fuel, escort, etc.) and the cost of the loss of human life.\footnote{We propose this without judgement on how to choose this function.}

\end{itemize}

For the sake of a clear definition of the problem, we make some additional assumptions:

\begin{itemize}

\item Only roads can be the targets of IEDs, not intersections or places of interest. This is done merely for clarification of the methods at hand, and is in no way a restriction of the capabilities of the modelling approach.

\item To ``block'' a road is to extend the time it would take to traverse it (as a clearing effort can be brought in) or to cause some level of damage to anyone trying to traverse it (by ignoring the clearing option); this investigation assumes that a road will not be traversed unless it is ``clear''.

\end{itemize}

We propose a general method for incorporating this problem into a family of methods based on the Generalized Linear Model, both in its static and dynamic forms, and by modifying concepts from the field of Complex Networks so that they can be incorporated into a GLM. This method is extremely flexible and allows for a large number of predictors and concepts to be added with minimal additional construction.

We begin in Section \ref{litreview} with a review of the literature from Operations Research that is relevant to the problem. We then introduce concepts in complex networks in Section \ref{complex-networks} that are relevant to the problem, namely centrality measures of nodes and edges, before introducing our synthesis of these ideas, Canadian Betweenness Centrality, in Section \ref{canadian-betweenness}. We show how to integrate these ideas with GLMs in Section \ref{ctp-in-glm} for the static case, before discussing dynamic extensions in Section \ref{dynamic-modelling}. 

We conclude with a discussion on the integration of expert knowledge in Section \ref{elicitation} before concluding with a short discussion on the improvements we believe can be made to this approach in Section \ref{future-extensions}.






\section{Literature Review\label{litreview}}

\subsection{Vehicle Routing in Transportation Research}

\citet{bertsimas1996ngvrrraau} provide an overview of vehicle routing methods from a deterministic view, extending these ideas to a stochastic or dynamic framework. They quote a canonical example: a utilities network is prone to failures that impede transmission; said failures vary in magnitude, timing and location according to some process. At the same time, maintenance units must make use of a transportation network to make repairs, so that the goal is to minimize the total system downtime by being efficiently deployed. 

\begin{figure}
\begin{center}
\includegraphics[width=10cm]{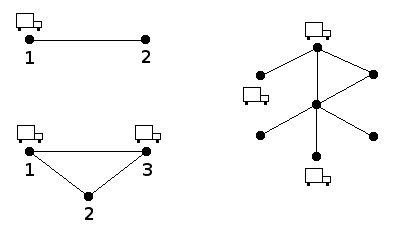}
\end{center}
\caption{Three vehicle routing scenarios. In each scenario, each location is prone to failures requiring the presence of a repair vehicle; these failures occur at stochastic intervals with associated costs. The problem to solve is the optimal placement of repair vehicles to minimize the costs incurred by failure. In the first, one vehicle is required to anticipate failures at two different locations; these scenarios become more complicated as additional locations and vehicles are added.\label{transport}}
\end{figure}

Figure \ref{transport} shows several examples of this class of problem. In the simplest case, the fleet consists of a single vehicle anticipating failure at one of two locations. As the number of locations and vehicles grows, ideal placement of the vehicles in the fleet requires the incorporation of several different elements: 

\begin{itemize}

\item The impact of several failures at once in the whole system, including their integrated behaviour. For example, in a communications system, two adjacent nodes suffering failure may be no worse than the failure of one of them, while the failure of two separate ``bridge'' nodes between two large groups may cripple all such contact.

\item The minimization of travel times of vehicles to respond to multiple simultaneous failures.

\item The connected impact of one failure on the likelihood of a failure in an adjacent node.

\end{itemize}

If the nodes are connected into a networked system, as in the case of a communications network, then the solution of this system's network properties (a stochastic problem) will be required to solve the deterministic vehicle deployment issue; for example, that a maintenance unit should be placed to respond to the nearest, most critical event that might occur. As a result, this does not necessarily require a joint solution of the two problems together. 

\citet{bertsimas1996ngvrrraau} refer to the broader area of dynamic transportation research. They include several subcategories relevant to our problem:

\begin{itemize}

\item Dynamic fleet management (\citet{powell1986smdvap} among others). A fleet of vehicles is distributed on the nodes of a network, ready to assist at other nodes as needed. This requires an algorithm to determine which vehicles should be dispatched at any given time to handle a service request given the likely distribution of later events. This application diverges from ours as we only consider a single source-destination pair at any one time. 

\item Dynamic traffic assignment (\citet{friesz1989dntacactocp}), in which individual units make decisions that optimize their own arrangements through traffic, mitigated by a decision-making process at each node along the way, such as minimizing a chosen cost function, as in this reference, according to a general Lagrangian equation for the system. This literature mostly examines the overall picture, in that different drivers make choices with implicit respect to one another. In this consideration, congestion is allowed to build on edges over time even if node traffic flow is allowed to be constant, as this is accounted for in the cost function. While this literature may have relevance to the general problem we tackle, as most of it is from the steady-state control perspective from a deterministic toolkit it is not likely to be directly adaptable. 

\item Dynamic shortest path problems (\citet{psaraftis1993dspanmac}) make up a very general class of algorithms. In the cited example, the ``dynamic'' aspect is that the properties of the nodes, rather than the edges, have stochastic properties. While the choice is likely made for the sake of keeping nodes as the independent unit, the properties of edges can be similarly modelled in our approaches. This literature tends to focus on the computational complexity of determining the shortest path, rather than an ensemble of short paths that may itself include the shortest path, which is consistent with our problem at hand.

\end{itemize}

Most of these dynamic problems are solved through linear or dynamic programming, allowing for a sizeable reduction in computing effort. It is not immediately clear how these methods can be adapted to systems where the variations on edge properties are mutually dependent, though the basic framework of solving simultaneously for all possible path outcomes is one that remains consistent in this framework.

\subsection{Stochastic Shortest Path Problems with Recourse}

A related problem to that under our investigation is the Canadian Traveller Problem, defined in \citet{andreatta1988sspr} and named by \citet{papadimitriou1991spwm} from the notion that roads may be closed due to stochastic intervention, namely a rode closure due to sudden snow blockage, where the realized state is only discovered when reaching one of the connecting nodes. The discovery of the shortest path is therefore determined dynamically as the system is traversed; the ``recourse'' nature of the problem is the technical term for dynamic rerouting during the traversal. This is the nature of the later work by \citet{polychronopoulos1996ssppr}. 

\begin{figure}
\begin{center}
\includegraphics[width=10cm]{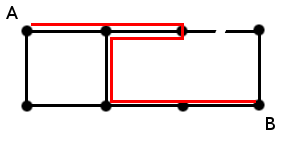}
\end{center}
\caption{A traveller on the graph from A to B discovers that an edge is untraversable only after arriving at one of its endpoints, and is forced to recalculate the optimal path based on this information. This is the general premise behind stochastic shortest path problems with recourse. \label{recourse}}
\end{figure}

This problem, as demonstrated in Figure \ref{recourse}, differs slightly from scenarios where an arc traversal time is stochastic but finite, as it may necessitate a reversal over a previously travelled arc for a finite solution to exist.

Among the investigations of this problem is \citet{karger2008eafctppat}, dealing with this problem by addressing scenarios for exact computational solutions to the problem when they are known to exist, such as directed acyclic graph structures; these cases do not necessarily translate to our current context.

\citet{bnaya2009ctprs} add the possibility of information gained by remote sensing; that is, the integration of non-local information gathering on the state of the system, under a simple information cost model. This approach may be integrated at some later stage with the aeriel detection systems proposed by \citet{royset2008oruasfiied} and discussed later in this review.

\citet{croucher1978nssp} deals with a somewhat related problem: the case when an acyclic graph is known but the path selection mechanism is stochastic. Namely, suppose that each arc $(i,j)$ emanating from node $i$ has distance $D_{ij}$ and traversal probability $p_{ij}$ (in the standard problem, $p_{ij}=1$ for all traversable arcs), and that there are $n_i$ outbound arcs from node $i$. Then given that it is decided to traverse arc $(i,j)$, the probability of traversing any of the other edges $(i,k)$ is $\frac{1-p_{ij}}{n_i}$. This scenario is considerably easier to solve through dynamic programming methods than others we have examined so far, but its applicability is less direct to the problem at hand.

Additionally, \citet{papadimitriou1991spwm} examine when the graph is completely unknown but embedded in a spatial manifold (such as a two-dimensional map) and the trajectory and distance to the goal is known, so that the goal is to produce a general strategy for traversing the graph that would minimize the total distance and/or cost to the traveller. 

This is a member of a more general class of problems in \textbf{reliability theory}. Rather than searching for single optimal solutions to shortest path problems, the purpose of a reliability study is to determine how a system responds to various failures. In this context, we seek to have estimates of the robustness of a system when certain paths become unavailable, or at a minimum more expensive.

On a very broad scale, \citet{banavar1999safetn} covers scaling relationships between the size of a networked system and the flow rates seen within. This notion of ``allometric scaling'' is a broad look at estimating travel time along a network given the size. Other examples of large-scale resilience estimation are listed in \citet{dorogovtsev2003en} though the focus is largely on grand-scale asymptotic results for classic families of network generation, namely the Erdos-Renyi-Gilbert random graph \citep{erdos1959rg, gilbert1959rg}, the Watts-Strogatz ``small world'' framework \citep{watts1998cdsn}, and the Barabasi-Albert ``scale-free'' construction \citep{barabasi1999esrn}. 

Appearing earlier in the literature is \citet{frank1969sppg}, covering a directly related problem: algorithms for calculating the {\emph distribution} of the shortest path in a graph if the edge lengths are stochastic (but finite) in nature. While only a rough guide to the process of solving the problem, this was clearly ahead of its time in thinking about this sort of problem in detail.

Finally, the Canadian Traveller Problem was also explored under the name ``bridge problem''  by \citet{blei1999spdud} due to the isomorphism between bridges-between-islands and roads-between-intersections.

\section{The Perspective of Complex Networks\label{complex-networks}}

The language of complex networks is well-suited to problems involving valued graphs such as a road system, where intersections can be seen as nodes and the roads are what connect them. Namely, a road network is considered to be a weighted graph $G(V,E)$, where a vertex $V_i$ represents an intersection between roads, typically a point of interest; the weight of an edge $E_{ij}$, between intersections $i$ and $j$, represents the distance between these points and the cost associated with traversing an edge in the problem of optimizing the total expense of a graph traversal. In general, we can represent a point of interest on a road as an intersection with degree 2 -- that is, a point along the road of interest that divides the road in two.

If there is a non-zero probability that this road is blocked, and that the blockage can only be discovered once we reach one end of the road, then we have the essence of a Canadian Traveller Problem, since the goal is to calculate the optimal route plan between two points, including all contingency plans. Before we introduce the CTP into the system, we first review how to model the graph of roads with IED deployments, as a subset of the greater road network, using the language of Generalized Linear Models.

The key to the approach that we will use is that we can model the probability that any particular stretch of road will have an IED deployment during a particular time interval, and that past events will be the key to this modelling. In particular, we assume that there are properties of the roads that lend themselves to deployments, both ``local'' in the sense of activities along the road itself, and ``global'' in the sense that the road holds importance as a connection between other places in the network.

\subsection{Generalized Linear Model Specifications for Deployment Probability Calculations}

Let $i,j \in \{1, ..., N\}$ index the nodes/intersections in the road system: $(i,j)$ then represents the direct road from $i$ to $j$ if it exists. Given that a road exists between two intersections (let an indicator $I_{ij}$ equal 1 if the road exists), the length of a road can be given as $L_{ij}$.

Considering a family of models for estimating the probability that a road has ever had an IED deployment using a Generalized Linear Model, namely a probit specification. This can be extended to other scenarios, such as a time-dependent structure in Section \ref{dynamic-modelling}, and incorporating the tactical position of the adversary as in \citet{singpurwalla2008nrde}; for this section, the use of the ``one-off'' specification is for illumination.

The general structure for the specification of a single road $(i,j)$, given that the road exists, is that a deployment was previously detected if $Y_{ij}=1$. If the probability of a previous deployment is $p_{ij}$, then $Y_{ij} \sim Be(p_{ij})$ so that

\begin{eqnarray}
\Phi^{-1}(p_{ij}) = & \mu & \mbox{ (baseline rate)} \\
+ & A_i & \mbox{ (rate given intersection i)} \\
+ & A_j & \mbox{ (rate given intersection j)} \\
+ & B_{ij} & \mbox{ (properties of edge (i,j))}.
\end{eqnarray}

It is the specification of each of these terms, including covariates on intersections and edges, that allows us to gauge their historical correlation with IED deployment likelihoods, and to lead to future predictions of deployment behaviour.

\subsubsection*{Intersection and Road Factors}

There are two basic categories of inputs at intersections. Let $X_i$ be a vector of properties of the intersection itself as a place of importance, such as proximity to a government building, school, mosque or other landmark of interest. This allows us to distinguish the globally defined properties of the intersection, which we label $Z_i$, which derive from the nature of the intersection in the traversal of the network itself. 

There may also be unobserved reasons for the domination of the intersection itself, which would suggest that a node-specific intercept term, either unpooled (each $\tau_i$ is determined independently, and all together sum to zero) or partially pooled ($\tau_i|\sigma \sim N(0,\sigma^2)$, and the best estimates for $\tau$ are parametrically shrunk toward zero), Depending on the frequency of past deployments, it may not be wise to include a node-specific intercept term in this equation, particularly if there are several intersections whose roads have never seen a deployment.

All together, these terms can be collected as

\[ A_i = X_i \alpha + Z_i \beta. \]

Each of these characteristics of intersections can also be present of the roads that lead to them. Let $X_{ij}$ identify a vector of local properties of the road itself (such as proximity to a local building of interest), and let $Z_{ij}$ identify those properties of the road that are due to the global nature of the road structure. Collecting these terms, we have

\[ B_{ij} = X_{ij}\gamma + Z_{ij}\delta. \]

We act as if the local properties of the road or intersection are known to the collector of the data, as in \citet{singpurwalla2008nrde}; in the next section we describe a number of possible global properties that measure the relative importances of nodes and edges in social network analysis that serve as the template for the importances of intersections and roads to transit.

\subsection{Including Traditional Centrality Measures As Predictors}

Social network analysis suggests two particular approaches to considering a node's importance to a network: closeness, or the ability to reach other nodes with a minimum of travel; and betweenness, or the importance of a node as it lies between the transit of two other nodes. While the latter is clearly the definition of most importance to the process on our network of interest -- the role that roads play is literally that of betweenness between two points of interest -- it is worth mentioning the role that closeness measures can play as an input for the nodal term.

\subsubsection*{Degree and Closeness}
The most basic form of closeness for a node is the number of other nodes with which it is in direct contact, which is known in a binary network as the ``degree'' of a node. In this case, it would represent the number of distinct roads that lead away from an intersection -- two at a point of interest along a road, three at a ``T'' junction, four where two roads cross, and so forth. One can include the degree of each intersection as a component of the $Z_i$ vector to check if a road attached to a particular intersection is more likely to have a deployment.

This concept is likely of lesser use than the idea of closeness centrality, a measure of the average distance between a node and all other nodes in the network system. The typical measure of distance in this case is the geodesic distance, or the shortest path from node $i$ to node $j$, symbolized as $d(i,j)$; the traditional measure of closeness centrality is then the inverse average distance from node $i$ to all others, as shown in \citet{freeman1979csncc},

\[ C_C(i) = \frac{n-1}{\sum_{k \neq i} d(i,k)}. \]

By taking the geodesic distance, we can of course assume that there is always available a shortest path of this length, and that this will always be the preferred path that any traveller will take. Modifications of the distance term can be made if necessary, so long as the distance between any two points remains finite, or that the term will include some sort of penalty for waiting for a repair of the road so that the journey can continue.

\subsubsection*{Betweenness}

The notion of betweenness stems from the importance of a node (or edge) in its placement for the transit between other nodes in the system. As disrupting this transition is one of the goals in IED deployment, adapting betweenness measures to the problem at hand is likely the most direct way of assessing the likelihood of a deployment.

Given a pair of nodes that serve as the destination and source (labelled $i$ and $j$ respectively), the standard geodesic betweenness of a node $k$ is measured in terms of all geodesic paths that connect $i$ to $j$ that contain $k$. Define $A_{ij}$ as the set of all unique paths between $i$ and $j$ with the geodesic traversal length $d(i,j)$; if $A_{ij}(k)$ is the set of all such paths that contain node $k$ as an intermediate step, then the ``betweenness'' of node $k$ with respect to path $(i,j)$ is the fraction of paths that contain it, $|A_{ij}(k)|/|A_{ij}|$. The overall betweenness measure of a node is then the average of this measure with respect to all pairs of nodes,

\[  C_B(k) = \frac{1}{(n-1)(n-2)} \sum_{i \neq k} \sum_{j \neq i,k} \frac{|A_{ij}(k)|}{|A_{ij}|}.  \]

This construction assumes that all pairs of nodes are equally important, and that a traveller will pick uniformly at random from all shortest paths, omitting any other paths that may be marginally longer. This does not mean that the construction can be adapted to allow for other eventualities.

There is an immediate adaptation to the importance of an edge, rather than a node, by replacing the node $k$ with the edge $(k,l)$; the betweenness of an edge is then reflected in the share of paths that traverse the edge $(k,l)$ when travelling from $i$ to $j$, defined as $A_{ij}(\{k,l\})$; the edge betweenness is then

\[ C_B(\{k,l\}) = \frac{1}{(n-1)(n-2)} \sum_{i \neq k,l} \sum_{j \neq i,k,l} \frac{|A_{ij}(\{k,l\})|}{|A_{ij}|}. \]

The removal of node $k$ or edge $\{k,l\}$ from the system is reflected in the betweenness statistic, but it does not in itself reflect the situation in the system after removal has itself happened. For the problem under consideration, the additional travel distance/time required in case the road is discovered to be blocked is a more apt measure of the road's importance, which we detail in the next section.

\section{Canadian Traveller Betweenness: How Much A Single Road Blockage Changes Travel Time\label{canadian-betweenness}}

We now introduce the essence of the Canadian Traveller Problem to this modelling approach. While the standard definitions of closeness and betweenness are deterministic in nature, the notion of Canadian Traveller Betweenness for a road/edge is an expected value for cases when an edge has a particular stochastic property, that the time to traverse it will be dependent on an uncertain event (the deployment of an IED) than can only be observed once one end of it has been reached. 

First, we review how optimal paths are found between any two points of interest in a transportation network when all road conditions are known using Dijkstra's algorithm, then show how this extends to the Canadian Traveller Problem specification.

After this review, we put forth the version of the problem that we find most compelling for thinking about this problem: assessing the importance of a road in terms of travel from a source to a destination. By considering the traveller problem focusing on one road at a time -- solving the problem with the road certain to be blocked (but unknown to the traveller), and with the road certain to be unblocked (again, unknown to the traveller) -- we assess the importance of the road to travel in the system itself, similar to the nature of betweenness centrality as explored in the previous section. This will then be integrated into the Generalized Linear Model approach in Section \ref{ctp-in-glm}.

\subsection{A Simple Example For A Single Source-Destination Pair\label{simple-example}}


\begin{figure}
\begin{center}
\includegraphics[width=0.5\linewidth]{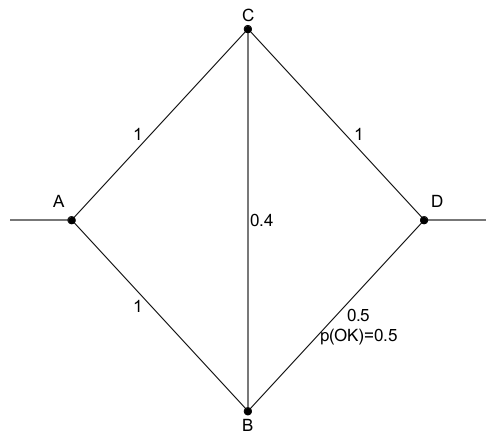}
\end{center}
\caption{When travelling from node A to node D, the traveller can choose the top route, with distance/cost 2, or try the bottom, which has 2 equally probable outcomes: the path BD is open, for a path $ABD$ and cost 1.5, or the path is closed, leading to a path $ABCD$ and cost 2.4. \label{simple-net}}
\end{figure}

To illustrate the challenges in this problem, consider Figure \ref{simple-net}, adapted from \citet{singpurwalla2008nrde}. The goal in this case is to travel from node A to node D; all edges except $BD$ are known to exist, while edge $BD$ may not exist with probability 0.5. Suppose that the goal is to find the travel plan that yields the minimum average travel time (noting that many other standards are also acceptable) and that the existence of $BD$ is only known upon arrival at $B$ or $D$.

A traveller taking the top path, $A$ to $C$ to $D$, makes a journey of distance 2. A traveller that tries the lower path will find a short road with probability 0.5, and a total journey $A-B-D$ with distance 1.5, or no direct road to $D$, forcing a detour back to $C$ and a journey $A-B-C-D$ with distance 2.6. Marginally, the expected length for the traveller choosing to try $B$ is 1.95, a little less than the traveller trying the certain route through $C$.

This construction serves to demonstrate the stepwise decision process that must be made by the traveller: a move toward the destination commits the traveller to a cost, but buys information about the landscape and reduces the total outcome space. Ahead of the actual transition along the graph, the user must assess the likelihood of each path being free or blocked before making a step in that direction, leading to a trade-off between ``discovery'' (the benefits of learning about the system) and ``progress'' (getting closer to the target in terms of the unblocked graph).

\subsection{Finding Optimal Paths and Travel Plans}

\subsubsection*{Dijkstra's Algorithm for Shortest Paths}
If the distance between all connections is known, the algorithm of \citet{dijkstra1959ntpcg} presents the optimal solution for the shortest path(s) from any one source node to all other nodes. The essence of the algorithm is as follows:

\begin{itemize}

\item Set the maximum shortest distance of all nodes, except the source node, to infinity; set the source to zero. Consider three classes of nodes: ``finished'', ``active'' and ``untested''; label the source node as ``active'' and the others as ``untested''.

\item For an ``active'' node, note all direct connections to ``untested'' nodes. For each connection, set the maximum distance of the untested node to the minimum of that and the current maximum distance of the active node plus the tie's distance.

\item Set the active node to ``finished''; set the untested node with the lowest maximum distance to ``active'', and repeat the procedure until all nodes are ``finished''.

\end{itemize}

This procedure gives the minimum distance from any other node to the source node. To find the shortest path, traverse the graph backwards by selecting the next node as that whose minimum distance equals the current node minus the connecting path length. 

\subsubsection*{The Cost of Direct Application}

One possible route to applying this methodology to the problem is through complete simulation. If each of the $r$ roads has only two possible states, then there are at most $2^r$ possible instantiations of the graph to check, and corresponding shortest paths can be solved for each solution. The challenge of the traveller -- and the essential breakdown of the problem -- is in deciding whether to take a single step along the most likely shortest path, or to take a more roundabout route in order to gather more information about the road system. 

Either of these methods induces a smaller Canadian Traveller Problem, where the destination remains the same, the source changes, and the uncertainty is lowered. In the end, this still requires the generation of a large number of possible realizations of the problem.

Barring the development of a complete solution to the problem, we are left with the same tools we had to begin with: to examine the shortest paths under each possibility, and to aggregate the probabilities of the respective scenarios in order to choose an optimal route plan, with a contingency at each step for whether the desired path remains optimal. The use of remote sensing, as proposed by \citet{bnaya2009ctprs}, suggests another factor to consider: whether a secondary mechanism can be used to check for the presence of a deployment ahead of, or parallel to, the main traveller, but this remains a hypothetical option with the same eventual constraints applying: multiple instances will have to be considered and integrated into the final solution.


\subsection{Defining Canadian Betweenness Centrality for an Edge\label{define-cbc}}

A road's importance to travel can be thought of in the sense of a potential trip length: how long the journey would be if the road were available, compared to the case when it is not available, when the discovery of availability can only be made when one of its endpoints is reached. 

This differs from the case when the graph layout is known ahead of time. Suppose there are multiple shortest paths between the source and destination, some but not all of which involve a road $(i,j)$. If the road state is known beforehand, the loss to the traveller on this path is zero, because the traveller can take one of the other paths and maintain the same travel distance. If the traveller picks all possible paths with some defined probability structure, in those cases where the path with $(i,j)$ is chosen will result in an increased travel distance.

Thus, define the Canadian Betweenness Centrality of an edge/road as the proportional expected increase in distance that a traveller would need make if it were removed, and that removal were only detected by the traveller when arriving at one of its endpoints.

Consider the network in Section \ref{simple-example}. To demonstrate how removing one of these five edges (and only one) would affect the average travel distance from $A$ to $D$, given that the average length of the preferred pathway on the lower track is 1.95:

\begin{itemize}

\item A removal of $AC$ would be discovered instantly, rather than in transition, and would cause the first move $A-B$. If the criterion is lowest average path length, only an irrelevant path would be removed; the increase in distance would be zero.

\item Removing $AB$ (also discovered instantly) would force the first move to be $A-C$. The traveller can then head straight to $D$ with distance 1, or head to $C$ on the 50\% chance that the road is open. If successful, the additional travel distance is 0.9; if not, the traveller must head back to $C$ before heading to $D$, for an additional distance of 1.8. The average path length in this scenario is 2.35, so that the traveller would be wise not to take it. The increase in distance $\Delta d_{AD}(A,C)$ is 0.4.

\item Removing $BC$ means that the failure of the bottom path would force the traveller to return to $A$ before trying the top path. With equal probability of lengths 0.9 and 4, and average length 2.45, the increase in distance $\Delta d_{AD}(B,C)$ is 0.5.

\item Removing $CD$ means that the only route to $D$ would be if the road were open. Letting the wait time for a road repair be $x_r$, there are two paths: the short path of length 0.9, and the path $A-B-C-B-(wait)-D$ of length $2.3+x_r$, for an average of $\Delta d_{AD}(C,D) = x_r/2-0.35$. 

\item If $BD$ is intact, the travelling distance is 0.9; if not, the distance is 2.4, for a total increase in distance of $\Delta d_{AD}(B,D) = 1.5$.

\end{itemize}

Noting that each of these terms was calculated with respect to the source-destination pair $(A,D)$, the same betweenness measure can be calculated for all pairs of nodes, and an average betweenness measure can be calculated for each edge in the system. The specification of the weights for each source-destination pair in the system can be chosen to be equal, or proportional to the number of trips taken per pair, or some other scheme chosen by the implementer. For equal weight, define the Canadian Betweenness Centrality as

\[ C_{CB}(i,j) = \frac{1}{(n-1)(n-2)} \sum_k \sum_{l < k} \Delta d_{kl}(i,j). \]

Given that the relative importance of each road is now calculated, each of these terms can be hypothetically included as a term in a model. This depends on the probability of a deployment being known so that a betweenness measure can be calculated with respect to all other ties in the system. In the next section we demonstrate how this can be accomplished through an iterative algorithm.

\section{Integrating CTP Estimates with GLM Constructions\label{ctp-in-glm}}

Now that we have a mechanism for describing the importance of a road to travel in a road network, we can include these terms in the GLM model for the likelihood of a deployment along each road. The only trick is that the deployment probabilities appear on both sides of the equation, as both the outcomes on the left and as components in the betweenness factor on the right. Here we describe an iterative method for solving for the importance of traveller betweenness in the deployment of an IED.

\begin{enumerate}

\item For each road $(i,j)$, choose a starting value for the probability of a deployment. For computation's sake, $p_{ij}=0$ is an appropriate (and fast) starting point. Hold this as $p_{ij}^{(0)}$.

\item Calculate the Canadian betweenness centrality $B_{ij}$ for each road in the system $(i,j)$ using the current deployment probability estimates $p_{ij}^{(0)}$.

\item Solve the linear model system

\[ Y_{ij} \sim Be(p_{ij}); \Phi^{-1}(p_{ij}) = \mu + (X_i + X_j)\alpha + (Z_i+Z_j) \beta + X_{ij}\gamma + Z_{ij}\delta \]

\noindent with one of the terms in the $Z_{ij}$ vector equal to $B_{ij}$.

\item With the estimates for $(\mu, \alpha, \beta, \gamma, \delta)$, calculate the new deployment probability $p_{ij}^{1}$.

\item Repeat steps 2-4 until the deployment probabilities have converged.

\end{enumerate}

As proposed, this algorithm is by no means fast; the Canadian Traveller Problem is \#P-complete and lacks a quick solution, making this algorithm easiest to run on small networks. For larger networks, rather than using the full space of possible graphs, one may instead sample from a subset of the $2^r$ possible graphs. This would be similar to the pseudolikelihood methods used to simulate from ERGMs \citep{crouch1998mcmcmlefpsnm}.

\subsection{Other Extensions To The One Time-Point Case}

This is a proposed road map for the integration of Generalized Linear Model methods for measuring and pooling information on IED deployment with the properties of the road system itself. There is a considerable number of possible improvements and developments that can be made on this framework.

\subsubsection*{Likely Travel Paths}

The relative importances of these roads have been determined on the assumption that the shortest (average) path would be preferable over any others. Realistically, there is also no guarantee that the traveller will take the shortest path, or that paths of slightly longer length will not be considered in a traveller's potential plans. Each of these importance measures can be adjusted to consider the utility of taking longer paths by chance, and estimating the additional travel caused by deployments on that basis.

\subsubsection*{Adversarial Deployment Patterns}

These decision methods have been made with the assumption that the deployment of IEDs is stochastic and exogenous in nature, not under the control of an adversary that can take the actions of the traveller into account other than the mathematical properties of the system. While there is certainly value in reducing the importance of a traveller's route choices down to local properties, it remains unclear how the change would be perceived by the adversary and how the next round of deployments would be affected.

\citet{singpurwalla2008nrde} introduced the notion of changing the likelihood function to reflect this, particularly in the notion that the lack of a past deployment makes a current deployment {\emph more} likely. The current specification is meant as a template for extending the likelihood function for past, present and future deployment mechanisms, since we specify the ingredients that will be introduced into the likelihood at each time point.


\section{Dynamic Modelling\label{dynamic-modelling}}

As outlined to this point, we have defined this method in terms of deployments during a single time interval; deployment probabilities were estimated {\em ex post facto} for each road in the system given their local and global properties. The use of the GLM framework, however, gives us a natural method to extend this approach to the dynamic time frame, and to allow new information to come into the method:

\begin{enumerate}

\item Extending the approach to include multiple-time-frame analysis in the GLM approach is straightforward. \citet{singpurwalla2008nrde} introduced the network routing problem through the specification of a time-dependent likelihood function, in the sense that a negative autocorrelation might be present---the lack of a previous deployment along a particular road may make a current deployment {\em more} likely to occur in the present frame (all else being equal). One possible scenario is to assume that more damage occurs on roads that have been through a lengthy repair process following a detonation.

\item Outside expert opinion can be introduced to the estimation process. There are at least three ways of doing this: by introducing expert opinion as a covariate in the model; by using these opinions as a mechanism for eliciting a prior distribution on the model parameters, and as a separate estimate that can be averaged with the model in some principled way.  

\end{enumerate}

This section details the addition of these characteristics into the single-step CTP-GLM method to produce a workable, dynamic method for improving the estimation process, as well as the predictive power of the model of adversarial behaviour given past actions.


The decision that needs to be made entails which route to take between two points on a map, given two or more possible paths that may be blocked by IED deployments. The decision-making process is sequential by nature: in the canonical Canadian Traveller Process, once an intersection is reached, the state of all its connected roads is known. This may be augmented by advance scouts, remote sensing or some other method with an associated cost, but in all senses there is still a lack of information that becomes a part of the decision process.


\subsection{Bayesian Updating Over Additional Time Periods\label{bayes-updating}}


One ultimate quantity of interest is $p(Y_{t+1}|Y_1,...,Y_t)$, the posterior predictive distribution of deployments at the next time point given past activity, after integrating out the model parameter estimates. These deployment probabilities, along with logistical considerations, will govern the route selection process. 

Recall the model for a single time step,

\[ Y_{ij} \sim Be(p_{ij}); \] 

\[ \Phi^{-1}(p_{ij}) = \mu + (X_i + X_j)\alpha + (Z_i+Z_j) \beta + X_{ij}\gamma + Z_{ij}\delta, \]

\noindent and consider a simple multistep version of the model: the deployment probabilities on each street, conditional on their characteristics and past histories, are independent and identically distributed. If this is the case, given a fixed number $N$ of time periods, the number of deployments on each road in the system can be modelled as a binomial distribution, with a simple stepwise updating scheme:

\begin{itemize}

\item Begin with the prior distribution for the model parameters $p(\mu, \alpha, \beta, \gamma, \delta)$.

\item Given the first time period of deployments, assemble the likelihood for the data, $p(Y_1|\mu, \alpha, \beta, \gamma, \delta)$, which under conditional independence breaks down as $\prod_{1 \leq i<j \leq n} p(Y_{ij,1}|\mu, \alpha, \beta, \gamma, \delta)$. Note that each of these components is a Bernoulli random variable.

\item Using this decomposition, solve for the posterior distribution after one time point, $p(\mu, \alpha, \beta, \gamma, \delta|Y_1)$, using whichever algorithm is desired -- Gibbs sampling, variational approximation, particle filtering -- making sure to adjust the estimate for the auxiliary variable set ${\widetilde{p_{ij}}}$, as it is needed for each Canadian betweenness $B_{ij}$ included in the $Z_I$ and $Z_{ij}$ terms.

\item Given this posterior distribution, repeat these steps for each new iteration of data to get the next posterior distribution $p(\mu, \alpha, \beta, \gamma, \delta|Y_1, Y_2)$, noting that conditional on the parameters $(\mu, \alpha, \beta, \gamma, \delta)$, the distribution for $Y_2$ does not depend on $Y_1$.

\end{itemize}

If the form is as simple as the binomial, this updating scheme is superfluous to the process, since we can typically conduct the entire operation in one step, from the prior distribution $p(\mu, \alpha, \beta, \gamma, \delta)$ directly to the posterior $p(\mu, \alpha, \beta, \gamma, \delta|Y_1,...,Y_N)$. There are also situations where the stepwise approximation will be lossy. However, there are cases when this updating scheme may prove useful, such as when the dependence structure is more complicated.

\subsection{Sample Time Dependence through Explicit Specification}

The previous method suggests the assumption that a deployment on a particular road would be independent of time and of other deployments on the same road at earlier times, an assumption that is quite likely untenable in real situations, since a recent deployment would probably discourage more of these in the immediate future (perhaps due to heightened vigilance, the effectiveness of deployment on a roadway under repair, and other reasons that may be explored by substantive experts.) \citet{singpurwalla2008nrde} suggests modifications to the likelihood function to incorporate this dependence directly; instead, we suggest that the explicit incorporation of previous observations would be a preferable way to introduce this dependence. 

One possible incorporation takes a Markovian form: the deployment in one time period depends explicitly on the previous period, as

\[ Y_{ij,t}|\mu, \alpha, \beta, \gamma, \delta,Y_{ij,t-1} \sim Be(p_{ij,t}); \]

\[ \Phi^{-1}(p_{ij}) = \tau Y_{ij,t-1} + \mu + (X_i + X_j)\alpha + (Z_i+Z_j) \beta + X_{ij}\gamma + Z_{ij}\delta, \]

\noindent so that for positive values of $\tau$, an attack the previous day would elevate the probability on the current day, and that this increase on probability would be identical for each road in the system (conditional on other observed characteristics.) A negative $\tau$ would correspond to a decrease in probability of an event if one had occurred in the previous time period.

In general, this can be extended to any number of past days as

\[ Y_{ij,t}|\mu, \alpha, \beta, \gamma, \delta,Y_{ij,t-1} \sim Be(p_{ij,t}); \]

\[ \Phi^{-1}(p_{ij}) = \sum_{k=1}^{K} \tau_k Y_{ij,t-k} + \mu + (X_i + X_j)\alpha + (Z_i+Z_j) \beta + X_{ij}\gamma + Z_{ij}\delta \]

\noindent to include any additional lags. For example, a series of negative values for each $\tau_k$, decreasing in magnitude as $k$ increases, would indicate that the further back one goes in time, the less a past deployment would affect the present, so that after a time, deployments would return to their apparent {\em status quo} rate.

\section{Elicitation of Expert Knowledge\label{elicitation}}

While the model-based approach can elucidate a good deal of information, a primary strength of the Bayesian approach is the ability to incorporate other information into the predictive framework. One method that makes the inclusion of expert opinion explicit is elicitation of prior distributions, (e.g. \citet{garthwaite2005smfepd}).  We suggest a simple method for converting from expert belief on roads into prior distributions on the parameters of interest on the model. A second approach is the addition of expert prediction as a covariate in the model, though this makes their uncertainty of their prediction harder to assess.  The third method we discuss is the some form of averaging of a parametric model prediction with properly calibrated expert opinion in the predictive phase of the model.

\subsection{Expert Loadings as Prior Weights}

Elicitation is a standard method for including expert information into both model and prior specification \citep{garthwaite2005smfepd}. By asking a series of questions of the expert, we can obtain information on the shape of the probability distribution that best describes their beliefs about likelihoods of events or strengths of associations.  We can then convert this information  into a distribution on the model parameters. Ideally, this information should be independent of anything known about the data under observation, such as the particular units of analysis (in this case, the roads themselves) but we can still adapt it to such circumstances when necessary.

If our goal is to learn something about the mechanisms in our model, such as the $(\alpha, \beta, \gamma, \delta)$ coefficients on local and global properties, then the wording of such questions may be difficult to elicit directly -- asking an expert about increased probabilities conditional on a covariate may be difficult to put into meaningful words, but asking an expert for their estimate of the probability of an IED deployment on a particular road in a time period is a tractable question. This forms the basis for our preliminary method for expert elicitation with respect to model parameters.

\begin{enumerate}

\item Select the expert from whom information will be elicited. Ask a series of ``warm-up'' questions to make the subject comfortable with probability assessments and uncertainties (see \citet{tversky1974juuhab} for an overview on these processes.)

\item For each road $(i,j)$ in the system, query the expert about their belief in the probability of a deployment of an IED in the time period in question, as well as their uncertainty about these probabilities. (See \citet{garthwaite2005smfepd} for more information on the process of elicitation.)

\item Set up the system of equations corresponding to

\[ \Phi^{-1}(p_{ij}) = \tau Y_{ij,t-1} + \mu + (X_i + X_j)\alpha + (Z_i+Z_j) \beta + X_{ij}\gamma + Z_{ij}\delta + \epsilon_{ij}, \]

\noindent and plug in the probability estimates for the $p_{ij}$; set $\epsilon_{ij} \sim N(0, \sigma^2)$, where $\sigma^2$ is an auxiliary variance parameter for this procedure. Solve for the ``posterior'' distribution of $(\mu, \alpha, \beta, \gamma, \delta|p_{ij})$ under this model beginning with a flat prior distribution.

\item Replace $p_{ij}$ with a draw from the distribution specified by the expert. Repeat the procedure.

\item Repeat the last step a large number of times until a series of distributions has been obtained. Take the average of these ``posteriors'' and label this as the elicited prior distribution for this expert.

\end{enumerate}

We think of this method as somewhat of a template that we can alter and refine in many different ways. Using this template, the principle of elicitation is firmly in place, using expert estimates to yield quantitative information for the model's prior parameters.  We can then use the procedure from Section \ref{bayes-updating}, starting with the elicited prior distribution as specified here.

\subsection{Covariate Addition}

Rather than consider  the expert opinion as the earlier basis for the model, we might instead treat the expert as a new source of information. If experts are  available at each point in the study under question, then their opinions on the probability of a deployment on any particular road can be added as covariates in the model, either as a prediction (0 or 1) or as a probability.

The disadvantages with this approach are obvious: it requires the continuous input of an expert during the process to be of any effect and it does allow the experts to calibrate their assessments against  one another and against the predictive strength of the model.  The direct interaction of these multiple sources of information may well affect the very estimates trying to be made, rather than treating the two sources as distinct (as has been observed in expert correlations of global warming estimates).

\subsection{Model Averaging and Linear Bayes Approaches}

Rather than worrying about directly fusing together two sources of information, an alternative is to teat the model's predictions and the expert's as two (or more) separate sources to be averaged together. This approach is common in prediction, and can be verified using repeated trials on the same experts, as above; sporting events (such as BCS championships in college football) and elections (such as the popular website FiveThirtyEight.com) are both frequently predicted using averages of expert and lay opinions plus more ``objective'' data models.  The weight applied to each predictor is adjusted with each successive time period or event of note, so that over time the predictions should improve in quality assuming all underlying assumptions remain true. 

Much empirical literature suggests that model averaging may be far from optimal when none of the predictors is based on a true model of the phenomenon under study (e.g. \citet{Geweke2010216}).  The alternative of using a pooled linear 
Bayesian predictor for this problem could benefit from careful exploration, such as the family of methods known as Bayesian Model Averaging \citep{raftery1995bmssr, raftery1997bmaflrm}.

\section{Some Additional Extensions\label{future-extensions}}

We have presented some standard extensions to a robust and flexible modeling approach that may be used for prediction in this kind of system; however, there is considerable room for the development and extension of these ideas.   We mention two examples of tasks as  obvious next  steps for this line of research.  

\subsection{Comprehensive Data Collection and Analysis}

While we could continue to develop these modeling ideas on simulated data sets and prototypical road systems, we cannot assess adversarial interest.  Nor can we  validate these modeling assumptions, without a substantially improved data, both for IED placement and the elicitation of experts.  Prototypical structures can prove to be useful to develop a concept, but certainly not to provide a meaningful illustration for policy purposes.

Once data exist for actual analysis of the proposed model, natural refinements and extensions may surface.

\subsection{Approximations to Canadian Betweenness}

The slowest part of the modelling process is the assessment of ``Canadian Betweenness'', which we showed in Section \ref{define-cbc} to be quite time-intensive to compute, making this impractical for larger networks without extensive pre-processing. If the measure appears to be a useful assessor of importance in real deployments, then an approximation to this measure may prove to be more useful in practice than bothering with the autoregressive form of the model that we are dealing with now, as well as being easier to handle the uncertainty in parameter estimates caused by the autoregressive components.

\bibliographystyle{ims}
\bibliography{actbib}

\end{document}